\documentclass[letterpaper, 10 pt, conference]{ieeeconf}
\usepackage{times}
\usepackage{color, colortbl, xcolor}
\usepackage{todonotes}

\usepackage{multicol}
\usepackage[bookmarks=true]{hyperref}
\usepackage{amsfonts}
\usepackage{amssymb}

\usepackage{algorithmic}
\usepackage[ruled,vlined,linesnumbered,titlenotnumbered]{algorithm2e} 
\usepackage{amsmath}
\usepackage{dsfont}
\usepackage{multicol}
\usepackage{mathtools}
\usepackage{subcaption}

\usepackage{balance}

\usepackage[%
    style=numeric-comp,
    sorting=none,
    backend=biber,
    sortcites=true,
    doi=false,
    firstinits=true,
    hyperref,
    isbn=false,
    eprint=false,
    block=none]
    {biblatex}
    
\renewbibmacro{in:}{}
\AtEveryBibitem{%
  \clearlist{language}%
  \clearfield{pages}%
}

\bibliography{references}

\newcommand{\numplayers}{N}
\newcommand{\cost}{J}
\newcommand{\runningcost}{\ell}
\newcommand{\horizon}{T}
\newcommand{\playerset}{\{1,\dots,\numplayers\}}
\newcommand{\feedback}{\gamma}
\newcommand{\feedbackset}{\Gamma}
\newcommand{\stepsize}{\eta}
\newcommand{\trustregion}{\epsilon}

\newcommand{\fpartial}[2]{\frac{\partial #1}{\partial #2}}
\newcommand{\fromlin}{\lambda}
\newcommand{\rdegree}{r}
\newcommand{\auxinput}{z}
\newcommand{\linx}{\xi}
\newcommand{\traj}{\mu}
\newcommand{\xdim}{n}
\newcommand{\udim}{k}

\newcommand{\mbb}[1]{\mathbb{#1}}

\def\atan2{\operatorname{atan2}}

\newtheorem{definition}{Definition}

\graphicspath{{./figures/}}    
\DeclareGraphicsExtensions{.pdf,.png}

\IEEEoverridecommandlockouts
\usepackage[textfont=md,font=footnotesize]{caption}

\usepackage{ifthen}
\newboolean{include-notes}
\setboolean{include-notes}{true}

\usepackage{lipsum}

\newcommand{\dfknote}[1]%
    {\textcolor{orange}{\textbf{[DFK: #1]}}}
\newcommand{\vrnote}[1]%
    {\textcolor{purple}{\textbf{[VR: #1]}}}

\newcommand{\remove}[1]%
    {\textcolor{red}{#1}}

\newcommand{\example}[1]%
{
\textbf{Running example:}
\textit{#1}
}

\title{\LARGE \bf An Iterative Quadratic Method for General-Sum Differential Games \\with Feedback Linearizable Dynamics}

\author{
David Fridovich-Keil*, Vicen\c{c} Rubies-Royo*, and Claire J. Tomlin
\thanks{
Department of EECS, UC Berkeley, \href{mailto:dfk@berkeley.edu}{\tt \small{\{dfk, vrubies, tomlin\}@eecs.berkeley.edu}}. $^*$ indicates equal contribution.}%
\thanks{This research is supported by an NSF CAREER award, the Air Force Office of Scientific Research (AFOSR), NSF's CPS FORCES and VeHICaL projects, the UC-Philippine-California Advanced Research Institute, the ONR BRC grant for Multibody Systems Analysis, a DARPA Assured Autonomy grant, and the SRC CONIX Center. D. Fridovich-Keil is also supported by an NSF Graduate Research Fellowship.}
}

\begin{document}

\maketitle
\thispagestyle{empty}
\pagestyle{empty}

\begin{abstract}
Iterative linear-quadratic (ILQ) methods are widely used in the nonlinear optimal control community. Recent work has applied similar methodology in the setting of multi-player general-sum differential games. Here, ILQ methods are capable of finding local equilibria in interactive motion planning problems in real-time.
As in most iterative procedures, however, this approach can be sensitive to initial conditions and hyperparameter choices, which can result in poor computational performance or even unsafe trajectories. 
In this paper, we focus our attention on a broad class of dynamical systems which are feedback linearizable, and exploit this structure to improve both algorithmic reliability and runtime.
We showcase our new algorithm in three distinct traffic scenarios, and observe that in practice our method converges significantly more often and more quickly than was possible without exploiting the feedback linearizable structure.
\end{abstract}

%\IEEEpeerreviewmaketitle
\section{Introduction}
\label{sec:intro}

In robotics, a wide variety of decision making problems, including low-level control, motion planning, and task planning, are often best expressed as optimal control problems.
Specific algorithms and solution strategies may differ depending upon factors such as system dynamics and cost structure; yet, modern methods such as model predictive control have proven extremely effective in many applications of interest.
Still, optimal control formulations are fundamentally limited to solving decision problems for \emph{a single agent}.

\begin{figure}[tbp]
    \centering
    \includegraphics[width=\linewidth, trim=70 180 70 210, clip=true]{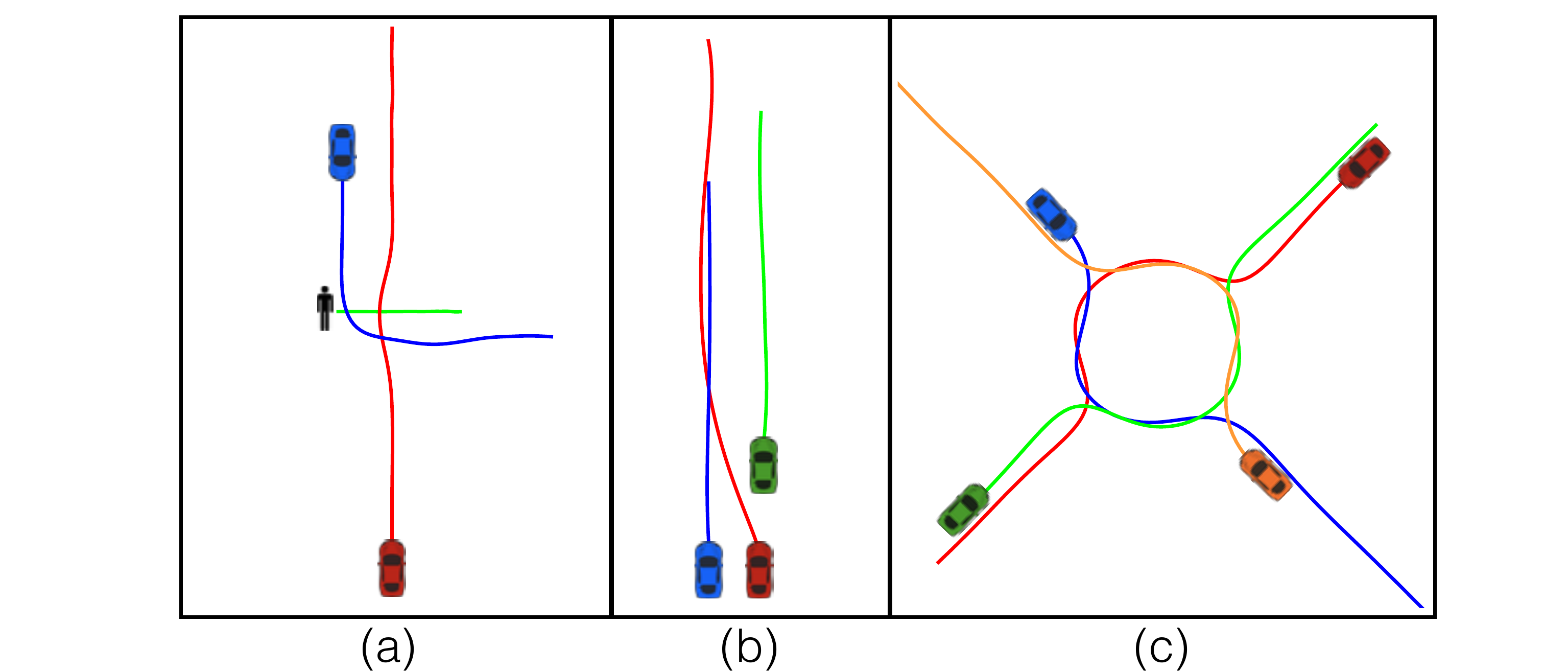}
    \caption{
    Three traffic scenarios---(a) intersection, (b) high speed overtaking, and (c) roundabout merging---which we formulate as differential games and use to benchmark the performance of our method and a baseline \cite{fridovich2019efficient}.
    }
    \label{fig:front}
    \vspace{-0.5cm}
\end{figure}

Dynamic game theory---the study of games played over time---provides a natural extension of optimal control to the multi-agent setting.
For example, nearby vehicles at an intersection (e.g., Fig.~\ref{fig:front}) mutually influence one another as they attempt to balance making forward progress in a desired direction while avoiding collision.
Abstractly, dynamic games provide each agent, or ``player,'' a separate input to the system, and allow each player to have a different cost function which they wish to optimize.
Players may wish to cooperate with one another in some situations and not cooperate in others, leading to complicated, coupled optimal play.
Moreover, different players may know different pieces of information at any point in time.
Optimal play depends strongly upon this information structure; a player with an informational advantage can often exploit that knowledge to the detriment of any competitors.
For example, in poker a player who cheats and looks at the top card in the deck is more certain of who will win the next hand, and hence can assume less risk in betting.

In this paper, we consider dynamic games played in continuous time, or differential games.
Historically, differential games were first studied in zero-sum (perfectly competitive) settings such as pursuit-evasion problems \cite{isaacs1951games}.
Here, optimal play is described by the Hamilton-Jacobi-Isaacs (HJI) PDE, in which the Hamiltonian includes a minimax optimization problem that encodes the instantaneous preferences of both players. 
These results extend to general-sum games as well, in which optimal play follows coupled HJ equations \cite{starr1969nonzero, starr1969further}.
Unfortunately, numerical solutions to these coupled PDEs often prove intractable because they operate on a densely discretized state space.
Approximate dynamic programming methods \cite{bertsekas1996neuro} offer a promising alternative; still, computational efficiency remains a significant challenge.

Recently, the robotics community has shown renewed interest in dynamic and differential games \cite{dreves2018generalized, dreves2019best, williams2018best}, with a variety of new approximate algorithms for identifying locally optimal play.
For example, Sadigh et al. \cite{sadigh2016planning} optimize the behavior of a self-driving car while accounting for the reaction of a human driver.
Wang et al. \cite{wang2019game} demonstrate a real-time iterative best response algorithm for planning competitive trajectories in a 6-player drone racing game.
Building upon the earlier sequential linear-quadratic method of \cite{mukai2000sequential} and the well-known iterative linear-quadratic regulator \cite{li2004iterative, jacobson1970differential}, our own prior work \cite{fridovich2019efficient} solves warm-started 3-player differential games in under 50 ms each, operating single-threaded on a consumer laptop.

In this paper, we extend and improve upon our previous work \cite{fridovich2019efficient} by exploiting the structure in a broad class of dynamical systems.
Many systems, including quadcopters and the planar unicycle and bicycle models commonly used to model automobiles, are feedback linearizable. 
That is, there exists a (nonlinear) control law which renders the closed-loop input-output dynamics of these nonlinear systems \emph{linear}.
Here, we develop an algorithm for identifying locally optimal play in differential games with feedback linearizable dynamics.
We establish theoretical equivalence between the solutions identified using this algorithm and those which do not exploit the feedback linearizable structure.
By exploiting the structure, however, our algorithm is able to take much larger steps at each iteration and generally converge to an equilibrium more quickly and more reliably than was previously possible.
Experimental results in Section~\ref{sec:results} confirm these computational advantages for the interactive traffic scenarios shown in  Fig.~\ref{fig:front}.

% One of the main goals of automation is to design systems capable of matching human performance on a wide array of tasks. In the field of robotics, optimal control theory is a particularly well suited mathematical framework for tackling this goal [refs], especially due to is its flexibility to integrate safety and robustness via differential games [refs]. Unfortunately, many optimal control approaches suffer from the `curse of dimensionality', hindering their use in real-time application. One recent approach \cite{fridovich2019efficient}

\section{Related Work}
\label{sec:related_work}

To put our work in context, here we provide a brief summary of iterative linear-quadratic (ILQ) methods of solving differential games, other approximate techniques of solving games, and common ways in which feedback linearization is used to accelerate motion planning.

\subsection{ILQ methods and other approximate techniques}
\label{subsec:ilqgames}

Iterative linear-quadratic (ILQ) methods are increasingly popular in the nonlinear model predictive control (MPC) and motion planning communities \cite{li2004iterative, van2014iterated}. These algorithms refine an initial control law at each iteration by forming a Jacobian linearization of system dynamics and a quadratic approximation of cost, and solving the resulting LQR subproblem. 
Because LQ games also offer an efficient solution, this approach has also been applied in the context of two-player zero-sum differential games by \cite{mukai2000sequential} and recently extended in \cite{fridovich2019efficient} to the $N$-player general-sum setting.
ILQ methods are \emph{local}. For optimal control problems, this means that they generally converge to local optima; for differential games, if they converge, they converge to local equilia.
Importantly, these methods scale favorably with state dimension (cubic), number of players (cubic), and time horizon (linear), and \cite{fridovich2019efficient} reports real-time operation for several three-player examples.

Iterative best response (IBR) algorithms comprise another class of methods for solving games. Here, in each iteration players sequentially solve (or approximately solve) the optimal control problem which results when all other players' strategies are fixed.
IBR has been demonstrated in a wide variety of settings, including congestion games \cite{jonsson2011scaling}, drone racing \cite{wang2019game}, and autonomous driving \cite{wangmingyu2019game}.
As in the case of ILQ methods, convergence is not generally guaranteed for arbitrary initializations. At best, IBR converges to local Nash equilibria (e.g., \cite{wang2019game}).
However, by reducing the game to a sequence of optimal control problems, IBR algorithms can take advantage of existing MPC and planning tools.

\subsection{Feedback linearization in motion planning}
\label{subsec:flin_planning}

Feedback linearization is a popular differential geometric control technique which renders a class of nonlinear systems' input-output response \emph{linear}.
We provide a brief technical overview of feedback linearization in Section~\ref{sec:feedback_linearization}; here, we summarize its relevance to motion planning.
One of the early successes of feedback linearization was its effectiveness in planning for chained systems, e.g., a car with multiple trailers \cite{murray1993nonholonomic, rouchon1993flatness}.
Feedback linearization (and the related notion of differential flatness) is also commonly used for minimum snap control of quadrotors \cite{mellinger2011minimum, richter2016polynomial}.
Here, the differentially flat structure of the underlying system dynamics allows planners to generate piecewise polynomial trajectories which the system can track exactly.
This concept is extended to the case of differential games in \cite{wang2019game}, where each iteration of IBR yields a new spline trajectory.

%\todoin{Look for feedback linearization + ILQR. Couldn't find it earlier, and if it does not exist in the literature would be nice to at least point out that our work easily includes that case.}

\section{Problem Formulation}
\label{sec:problem}

%\todoin{Explain what a differential game is and that we'll assume control affine dynamics, feedback linearizable, etc.}

We consider $\numplayers$-player, general-sum differential games with control-affine dynamics. That is, we presume that the game state $x \in \mbb{R}^{\xdim}$ evolves as
\begin{align}
\label{eqn:dynamics}
\dot x = f(x) + \sum_{i = 1}^\numplayers g_i(x) u_i~,
\end{align}
where $u_i \in \mbb{R}^{\udim_i}$ is the control input of player $i$. In our examples (Section~\ref{sec:results}), $x$ will be the concatenated states of multiple subsystems, but this is not strictly necessary.
We assume that \eqref{eqn:dynamics} is full-state feedback linearizable, i.e. there exist outputs $y = h(x)$ such that $y$ and finitely many of its time derivatives evolve linearly as a function of some auxiliary inputs $\auxinput_i \in \mbb{R}^{\udim_i}$, for some control law $u_i := u_i(x, \auxinput_i)$. A brief review of feedback linearization may be found in Section~\ref{sec:feedback_linearization}.

Next, we suppose that each player $i$ wishes to minimize a running cost $\runningcost_i$ over finite time horizon $\horizon$:
\begin{align}
\label{eqn:cost}
\cost_i(u_1, \dots, u_\numplayers) := \int_0^T \runningcost_i(t, x, u_1, \dots, u_\numplayers) dt~.
\end{align}
We shall require $\ell_i$ to be $C^2$ in $x, u_j, \forall j$, uniformly in time $t$. Player $i$'s total cost $\cost_i$ then depends explicitly upon each player's control input signal $u_i(\cdot)$ and implicitly upon the initial condition $x(0)$.

Finally, we presume that each player $i$ has access to the state $x$ at every time $t$, but \emph{not} other players' control inputs $u_j, j\ne i$, i.e.
\begin{align}
\label{eqn:information_pattern}
u_i(t) \equiv \feedback_i\big(t, x(t)\big)
\end{align}
for some measurable function $\feedback_i : [0, \horizon] \times \mbb{R}^\xdim \to \mbb{R}^{\udim_i}$. We shall denote the set of such functions $\feedbackset_i$. For clarity, we shall also overload the notation of costs $\cost_i(\feedback_1; \dots; \feedback_\numplayers) \equiv \cost_i\big(\feedback_1(\cdot, x(\cdot)), \dots, \feedback_\numplayers(\cdot, x(\cdot))\big)$.

Thus equipped with dynamics \eqref{eqn:dynamics}, costs \eqref{eqn:cost}, and information pattern \eqref{eqn:information_pattern}, in principle, we seek Nash equilibria of the game.

\begin{definition} (Nash equilibrium, \cite[Chapter 6]{basar1999dynamic})
A set of strategies $(\feedback_1^*, \dots, \feedback_\numplayers^*)$ constitute a Nash equilibrium if no player has a unilateral incentive to deviate from his or her strategy. Precisely, the following inequality must hold for each player $i$:
\begin{align*}
\cost_i^* &\equiv \cost_i(\feedback_1^*, \dots, \feedback_{i-1}^*, \feedback_i^*, \feedback_{i+1}^*, \dots, \feedback_\numplayers^*) \\
&\le \cost_i(\feedback_1^*, \dots, \feedback_{i-1}^*, \feedback_i, \feedback_{i+1}^*, \dots, \feedback_\numplayers^*), \forall \feedback_i \in \feedbackset_i~.
\end{align*}
\end{definition}

In practice, we may only be able to check if these conditions are satisfied locally in the neighborhood of strategy $\feedback_i^*$, i.e., find \emph{local} Nash equilibrium. 
As noted in \cite{fridovich2019efficient} ILQ methods for games are known \emph{not} to find even local Nash equilibria, but still be competitive and, importantly, efficient to compute.

\section{Background: Feedback Linearization}
\label{sec:feedback_linearization}

% For the remainder of this paper we will assume the dynamics of all agents to be control-affine

% \begin{equation}
%     \label{eq:control-affine}
%     \begin{aligned}
%     &\dx = f(x) + g(x)u\\
%     &y = h(x)
%     \end{aligned}
% \end{equation}
% with state $x \in \mathbb{R}^n$, control input $u \in \mathbb{R}^m$ and output $y \in \mathbb{R}^m$.
This section provides a brief review of feedback linearization, a geometric control technique popularly used across a wide range of robotic applications including manipulation, quadrotor flight, and autonomous driving.

Recall dynamics \eqref{eqn:dynamics}, and define the matrix $g(x) := [g_1(x), \dots, g_\numplayers(x)] \in \mbb{R}^{\xdim \times \udim}$ and vector $u^T = [u_1^T, \dots, u_\numplayers^T] \in \mbb{R}^{\udim}$, with $\udim = \sum_i \udim_i$ the total control dimension. Thus, \eqref{eqn:dynamics} may be rewritten as
\begin{equation}
\label{eqn:flin_dyn}
\begin{aligned}
\dot x = f(x) + g(x) u,~y = h(x)
\end{aligned}
\end{equation}
where $y$ is the output of the system, and the functions $f, g$, and $h$ are sufficiently smooth.

\subsection{Mechanics}
\label{subsec:flin_mechanics}

Suppose that \eqref{eqn:flin_dyn} has well-defined vector relative degree $(\rdegree_1, \dots, \rdegree_\udim)$ \cite[Definition 9.15]{sastry1999nonlinear} and is full-state feedback linearizable. Then, there exists a matrix $M(x)$ and vector $m(x)$ such that the time derivatives of the outputs $y$ follow
% In this section we consider the more general case where $y,u \in \mathbb{R}^\udim, \udim > 1$ and $1 \le \numplayers \le \udim$. Analogously to the previous section, one can differentiate each entry $y_i$ of the output vector with respect to time $\rdegree_i$ times and obtain the generalization of \eqref{eq:siso-flin-gamma}
\begin{equation}
    \label{eq:mimo-flin-gamma}
    \begin{aligned}
    [y_1^{(\rdegree_1)},\dots,y_\udim^{(\rdegree_\udim)}]^T = M(x)u + m(x). 
    \end{aligned}
\end{equation}

Presuming the invertibility of the so-called ``decoupling matrix'' $M(x)$, we may design the following feedback linearizing control law as a function of both state $x$ and an \emph{auxiliary input} $\auxinput\in\mbb{R}^\udim$:
% Unlike \eqref{eq:siso-flin-gamma}, the requirement for the control law to be well defined will be for $M(x): \mathbb{R}^\xdim \to \mathbb{R}^{\udim \times \udim}$ to be invertible for all $x \in \mathcal{S}$. $M(x)$  is known as the \emph{decoupling matrix}, and any point $x$ at which $M(x)$ drops rank is known as a \emph{singularity}. The term $m(x)$ will be denoted as the \emph{drift term}. Presuming $M(x)$ invertible, we have the following feedback linearizing control law
\begin{equation}
    \label{eq:flin-mimo-control-law}
    \begin{aligned}
    u(x,\auxinput) = M^{-1}(x)(\auxinput - m(x)),
    \end{aligned}
\end{equation}
which renders the input-output dynamics linear in the new auxiliary inputs $\auxinput$:
\begin{align}
\label{eq:flin-linsys}
[y_1^{(\rdegree_1)},\dots,y_\udim^{(\rdegree_\udim)}]^T = \auxinput.
\end{align}

Note that, as for $u$ in \eqref{eqn:dynamics} we shall consider $\auxinput^T = [\auxinput_1^T, \dots, \auxinput_\numplayers^T]$ to be a concatenation of auxiliary inputs for each player, with $\auxinput_i \in \mbb{R}^{\udim_i}$.

\subsection{Change of coordinates}
\label{subsec:change_coordinates}

We have seen how a careful choice of feedback linearizing controller $u(x, \auxinput)$ renders the dynamics of the output $y$ and its derivatives linear. 
Define the state of this linear system as $\linx := [y_1, \dots, y_1^{(\rdegree_1 - 1)}, \dots, y_\udim, \dots, y_\udim^{(\rdegree_\udim - 1)}]^T$.
Just as there is a bijective map \eqref{eq:flin-mimo-control-law} between control $u$ and auxiliary input $\auxinput$ whenever $M(x)$ is invertible, there is also a bijection between state $x$ and linear system state $\linx, x = \fromlin(\linx)$ \cite{sastry1999nonlinear} because \eqref{eqn:flin_dyn} is full-state feedback linearizable. We shall use both bijective maps (and their derivatives) in Section~\ref{sec:methods} to rewrite costs \eqref{eqn:cost} in terms of the linearized dynamics \eqref{eq:flin-linsys}.

%Feedback linearization produces control law \eqref{eq:flin-mimo-control-law} which results in a \textit{linear} dynamical system of the form $[y_1^{(\gamma_1)},...,y_m^{(\gamma_m)}]^T = v$. This allows the control of the system's outputs via LQR and other similar tools from linear systems theory. This new dynamical system implicitly defines a new state vector $\xi := [y_1^{(\gamma_1-1)},...,y_m^{(\gamma_m-1)}]^T$ in terms of the outputs (and their time derivatives) of the original non-linear system. In general, assuming full observability of the non-linear system states, there also exists a bijective map $z(x): \mathbb{R}^m \to \mathbb{R}^m$ mapping $\xi$ to $x$. Therefore, the running cost $\runningcost_i(t, x, u_1, \dots, u_\numplayers)$ can also be transformed into a cost in terms of $\xi$ and $v$ of the form $\hat\runningcost_i = \runningcost_i(t, z(\xi), u_1(z(\xi),v), \dots, u_\numplayers(z(\xi),v))$. Importantly, this cost $\hat\runningcost_i$ will have a well-defined analytical expression, and, thus, an equally well-defined gradient and hessian.   

\section{Methods}
\label{sec:methods}

In this section we present our main contribution, a computationally stable and efficient algorithm for identifying local equilibria of general-sum games with feedback linearizable dynamics.
We begin in Section~\ref{subsec:example_flin} by computing a feedback linearizing controller for unicycle dynamics, which we shall use as a running example throughout the paper.
Then, in Section~\ref{subsec:transforming_costs} we show how to transform the costs for each player to depend upon linear system state $\linx$ and auxiliary inputs $\auxinput_i$ rather than state $x$ and controls $u_i$.
In Section~\ref{subsec:main_alg} we introduce the main algorithm, and finally in Section~\ref{subsec:soundness} we summarize the effects of using feedback linearization.

\subsection{Feedback linearization by example}
\label{subsec:example_flin}

Consider the following (single player) 4D unicycle dynamical model:
\begin{align}
\label{eqn:unicycle_dyn}
\dot x = 
\begin{bmatrix}
\dot p_x\\
\dot p_y\\
\dot \theta\\
\dot v
\end{bmatrix} = \begin{bmatrix}
v \cos \theta\\
v \sin \theta\\
w \\
a
\end{bmatrix},~y = \begin{bmatrix}
p_x\\
p_y
\end{bmatrix}
\end{align}
representing the evolution of the positions $p_x$ and $p_y$, the orientation $\theta$, and speed $v$. The inputs $w$ and $a$ represent the angular rate and the acceleration. By taking time derivatives of the output $y$ following the procedure from Section~\ref{sec:feedback_linearization}, we obtain the new set of states ${\linx = [p_x, \dot{p}_x, p_y, \dot{p}_y]^T}$ for the linearized system. Differentiation reveals that
\begin{equation}
  \begin{bmatrix}
    \ddot{p}_x\\
    \ddot{p}_y\\
 \end{bmatrix} =
 \begin{bmatrix}
    -v \sin \theta & \cos \theta\\
     v \cos \theta  & \sin \theta\\
 \end{bmatrix} 
  \begin{bmatrix}
    w\\
    a\\
 \end{bmatrix}.% = M(x)u + m(x)
\end{equation}
From this result, we compute the inverse decoupling matrix and drift term as
\begin{align}
\label{eqn:unicycle_decoupling}
    M^{-1}(x) = 
    \begin{bmatrix}
   -\sin \theta / v & \cos \theta / v\\
    \cos \theta & \sin \theta \\
    \end{bmatrix},~
    m(x) = 0.
\end{align}
Finally, we can also explicitly derive the state conversion map $\fromlin(\linx)$
\begin{align}
    \fromlin(\linx) = 
    \begin{bmatrix}
    p_x \\
    p_y\\
    \sqrt{\dot{p}_x^2 + \dot{p}_y^2}\\
    \tan^{-1}(\frac{\dot{p}_y}{\dot{p}_x})
    \end{bmatrix}.
\end{align}

Now, consider a differential game with two players, each of whom independently follows dynamics \eqref{eqn:unicycle_dyn}. The inverse decoupling matrix $M^{-1}(x)$ and the Jacobian of the state conversion map $\fromlin$ for the full system will be block diagonal.

\subsection{Transforming costs}
\label{subsec:transforming_costs}

So far, we have introduced feedback linearization and shown how to derive the mappings from auxiliary input $\auxinput$ to control $u$ and linearized system state $\linx$ to state $x$. 
To exploit the feedback linearizable structure of \eqref{eqn:flin_dyn} when solving the game, we must rewrite running costs $\runningcost_i(t, x, u_1, \dots, u_\numplayers)$ in terms of $\linx$ and  $\auxinput$.
Overloading notation, we shall denote the transformed running costs as
\begin{equation}
\label{eqn:transformed_cost}
\begin{aligned}
\runningcost_i(t;\,&\linx; \auxinput_1; \dots; \auxinput_\numplayers) \equiv\\
&\runningcost_i\Big(t; \fromlin(\linx); u_1\big(\fromlin(\linx), \auxinput_1\big); \dots; u_\numplayers\big(\fromlin(\linx), \auxinput_\numplayers\big)\Big),
\end{aligned}
\end{equation}
where $u_i(\fromlin(\linx), \auxinput_i)$ is given in \eqref{eq:flin-mimo-control-law}.

Section~\ref{subsec:main_alg} presents our main algorithm; a core step will be to compute first and second derivatives of each player's running cost with respect to the new state $\linx$ and inputs $\auxinput_i$. 
This may be done efficiently using the chain rule and exploiting known sparsity patterns for particular systems and costs.
For completeness, however, we shall ignore sparsity and illustrate computing the first derivative of $\runningcost_i$ with respect to the $j^\text{th}$ dimension of $\linx$, denoted $\linx_j$:
\begin{align}
\label{eqn:dldxi}
\fpartial{\runningcost_i}{\linx_j} = \sum_{p=1}^\xdim \fpartial{\runningcost_i}{x_p} \fpartial{x_p}{\linx_j} + \sum_{n=1}^\numplayers \sum_{p=1}^{\udim_n} \fpartial{\runningcost_i}{u_{n, p}}  \sum_{q=1}^\xdim \fpartial{u_{n, p}}{x_q} \fpartial{x_q}{\linx_j}
\end{align}
where $u_{n,p}$ is the $p^{\text{th}}$ entry of the $n^{\text{th}}$ player's control input.

Second derivatives may be computed similarly, though again we stress that for specific dynamics and cost functions it is often much more efficient to exploit the \emph{a priori} known sparsity of partial derivatives. Interestingly, we also observe that the terms arising from the second sum in \eqref{eqn:dldxi}, which account for the state-dependence of the feedback linearizing controllers \eqref{eq:flin-mimo-control-law}, are often negligible in practice and may be dropped without significant impact on solution quality. %\dfknote{add more on hessians for contorl cost}

\subsection{Core algorithm}
\label{subsec:main_alg}

\begin{algorithm}[t]
\KwIn{initial linearized system state $\linx(0)$ and control strategies $\{\feedback_i^0\}_{i \in \playerset}$, time horizon $\horizon$}
\KwOut{converged control strategies $\{\feedback_i^*\}_{i \in \playerset}$ for the linearized system}

\For{iteration $p = 1, 2, \dots$}{
    $\traj^p \equiv \{\hat \linx(t), \hat \auxinput_i(t)\}_{i \in \playerset, t \in [0, \horizon]} \leftarrow$\\ \Indp getTrajectory$\big(\linx(0), \{\feedback_i^{p-1}\}\big)$\nllabel{line:operating_pt}\;
    \Indm
    $\{l_i(t), Q_i(t), R_{ij}(t)\} \leftarrow$ quadraticizeCost$\big(\traj^p\big)$\nllabel{line:quadraticize}\;
    $\{\tilde \feedback_i^p\} \leftarrow \textnormal{solveLQGame}\big(\{l_i(t), Q_i(t), R_{ij}(t)\}\big)$\nllabel{line:solve_lq}\;
    $\{\feedback_i^p\} \leftarrow \textnormal{stepToward}\big(\{\feedback_i^{p-1}, \tilde \feedback_i^p\}\big)$\nllabel{line:step}\;
    \If{converged\nllabel{line:converged}}{
        \Return{$\{\feedback_i^p\}$}
    }
}
\caption{Feedback Linearized Iterative LQ Games \label{alg:flin_ilq_game}}
\end{algorithm}

Like the original iterative LQ game algorithm, we proceed from a set of initial strategies $\feedback_i$ for each player---understood now to map from $(t, \linx)$ to $\auxinput_i$---and iteratively refine them by solving LQ approximations.
Our main contribution, therefore, lies in the transformation of the game itself into the coordinates $\linx, \auxinput_i$ which correspond to feedback linearized dynamics.
As we shall see in Section~\ref{sec:results}, iterative LQ approximations are much more stable in the transformed coordinates and converge at least as quickly.

Algorithm~\ref{alg:flin_ilq_game} outlines the major steps in the resulting algorithm. We begin at the given initial condition $\linx(0)$ for the linearized system and strategies $\feedback_i^0$ for each player.
Note that these strategies define control laws \emph{for the linearized system}, i.e. $\auxinput_i(t) \equiv \feedback_i\big(t, \linx(t)\big)$.

At each iteration, we first (Algorithm~\ref{alg:flin_ilq_game}, line~\ref{line:operating_pt}) integrate the linearized dynamics \eqref{eq:flin-linsys} forward to obtain the current operating point $(\hat\linx(\cdot), \{\hat\auxinput_i(\cdot)\})$. Then (Algorithm~\ref{alg:flin_ilq_game}, line~\ref{line:quadraticize}), we compute a quadratic approximation to each player's running cost in terms of the variations $\delta \linx := \linx - \hat\linx$ and $\delta \auxinput_j := \auxinput_j - \hat \auxinput_j$
\begin{equation}
\label{eqn:quadraticize}
\begin{aligned}
\runningcost_i(&t;\linx; \auxinput_1; \dots; \auxinput_\numplayers) - \runningcost_i(t; \hat\linx; \hat\auxinput_1; \dots; \hat\auxinput_\numplayers) \approx \delta\linx^Tl_i(t) +\\
&\frac{1}{2}\delta\linx^TQ_i(t) \delta\linx + \frac{1}{2}\sum_{j=1}^\numplayers \delta\auxinput_j^T(R_{ij}(t) \delta\auxinput_j + 2r_{ij}(t)),
\end{aligned}
\end{equation}
using the chain rule as in \eqref{eqn:dldxi} to compute the terms $l_i, Q_i$ and $R_{ij}$ for each player.

Equipped with linear dynamics \eqref{eq:flin-linsys} and quadratic costs \eqref{eqn:quadraticize}, the solution of the resulting general-sum LQ game is given by a set of coupled Riccati differential equations, which may be derived from the first order necessary conditions of optimality for each player \cite[Chapter 6]{basar1999dynamic}. 
In practice (Algorithm~\ref{alg:flin_ilq_game}, line~\ref{line:solve_lq}), we numerically solve these equations in discrete-time using a time step of $\Delta t$.
If a solution exists at the $p^{\text{th}}$ iteration, it is known to take the form
\begin{align}
\label{eqn:affine_feedback}
\tilde\feedback_i^p(t, \linx) \equiv \hat\auxinput_i(t) - P_i^p(t) \big(\linx(t) - \hat\linx(t)\big) - \alpha_i^p(t)
\end{align}
for matrix $P_i^p(t)$ and vector $\alpha_i^p(t)$ \cite[Corollary 6.1]{basar1999dynamic}.

We cannot simply use these strategies at the $(p+1)^{\text{th}}$ iteration or we risk diverging, however, without further assumptions on the curvature and convexity of running costs $\runningcost_i$. In fact, these costs are generally nonconvex when expressed in terms of $\linx$ and $\auxinput_j$ \eqref{eqn:transformed_cost}, which necessitates some care in updating strategies.
To address this issue (Algorithm~\ref{alg:flin_ilq_game}, line~\ref{line:step}), we follow a common practice in the ILQR and sequential quadratic programming literature (e.g., \cite{tassa2014control}) and introduce a step size parameter $\stepsize \in (0, 1]$:
\begin{align}
\label{eqn:stepsize}
\feedback_i^p(t, \linx) = \hat\auxinput_i(t) - P_i^p(t) \big(\linx(t) - \hat\linx(t)\big) - \eta \alpha_i^p(t).
\end{align}
Observe that, taking $\stepsize = 0$ and recalling that $\linx(0) = \hat \linx(0)$, we recover the previous open-loop control signal $\feedback_i^p(t, \linx) = \hat\auxinput_i, \forall t \in [0, \horizon]$. Taking $\stepsize = 1$, we recover the LQ solution from this iteration \eqref{eqn:affine_feedback}.
As is common in the literature, we perform a backtracking linesearch on $\stepsize$, starting with initial value $\stepsize_0$ and terminating when the trajectory that results from \eqref{eqn:stepsize} satisfies a trust region constraint at level $\trustregion$. In our experiments, we use an $L_{\infty}$ constraint, i.e. \begin{align}
\label{eqn:trust_region}
    \|\linx(t) - \hat \linx(t)\|_{\infty} < \trustregion, \forall t,
\end{align}
and check that $M^{-1}$ exists at each time.% (see \ref{subsec:soundness} remark \ref{remark:drawbacks}). 

\subsection{Effect of feedback linearization}
\label{subsec:soundness}

%\begin{remark} (Criterion for Convergence to Local Nash Equilibrium)
%\label{remark:criterion}
%Suppose that Algorithm~\ref{alg:flin_ilq_game} converges to strategies $\{\feedback_i^*\}$. Then, from \cite[Theorem 1]{fridovich2019efficient} and presuming the invertibility of $M$ we have that if Hessians $D_{xx}\big(\runningcost_i(t)\big), D_{u_i u_j}\big(\runningcost_i(t)\big) \succeq 0, \forall t$ at convergence, then the \emph{open-loop} controllers defined by $\auxinput_i(t) = \hat \auxinput_i^*(t)$ comprise a local Nash equilibrium in open-loop strategies for the original system. That is, taking
%$$u^*(t) = M^{-1}\Big(\fromlin\big(\hat\linx^*(t)\big)\Big)\Big(\hat\auxinput^*(t) - m\big(\fromlin(\hat\linx^*(t))\big)\Big)$$
%we obtain a local Nash equilibrium in open loop strategies for the game defined by dynamics \eqref{eqn:dynamics} and costs \eqref{eqn:cost}.
%end{remark}

% \begin{remark} (Benefits of Feedback Linearization)
% \label{remark:benefits}
In comparison to the non-feedback linearizable case, the linearized dynamics \eqref{eq:flin-linsys} are independent of trajectory (and hence also of iteration). That is, in the non-feedback linearizable case \cite{fridovich2019efficient}, each iteration begins by constructing a Jacobian linearization of dynamics \eqref{eqn:dynamics}; this is superfluous in our case. As a consequence, large changes in auxiliary input $\auxinput$ between iterations---which lead to large changes in state trajectory---are trivially consistent with the feedback linearized dynamics \eqref{eq:flin-linsys}. By contrast, a large change in control $u$ may take the nonlinear dynamics \eqref{eqn:dynamics} far away from the previous Jacobian linearization, which causes the algorithm from \cite{fridovich2019efficient} to be fairly sensitive to step size $\stepsize$ and trust region size $\epsilon$. We study this sensitivity more carefully in Section~\ref{sec:results}.
% \end{remark}

% \begin{remark} (Drawbacks of Feedback Linearization)
% \label{remark:drawbacks}
Finally, it is important to note that while many systems of interest (e.g., manipulators, cars, and quadrotors) are feedback linearizable, this is not true of all systems.
Additionally, there are two drawbacks of our algorithm that deserve mention. First, we must take care to avoid regions in which $M^{-1}$ does not exist. We accomplish this by designing costs that penalize proximity to singularities. While this can potentially limit the range of behaviors, many motion problems naturally incorporate these costs. 
Second, the transformed costs $\runningcost_i(t; \linx; \dots)$ may have much more varied, extreme curvature than the original costs $\runningcost_i(t, x, \dots)$.
In some cases, this can make Algorithm~\ref{alg:flin_ilq_game} sensitive to linesearch parameters $\eta_0$ and $\trustregion$, even offsetting the benefits mentioned above.
We defer further discussion and empirical study for Section~\ref{sec:results}.
%rewriting running costs $\runningcost$ in terms of linearized system state $\linx$ and auxiliary input $\auxinput$ can be algebraically intensive, and at times the resulting computational complexity (even to evaluate an analytical expression) is noticeable.
% \end{remark}

%\input{details.tex}

\section{Results}
\label{sec:results}

In this section, we study the empirical performance of Algorithm~\ref{alg:flin_ilq_game}.
In Section~\ref{subsec:stability}, we quantify the improvements in algorithmic stability from Section~\ref{subsec:soundness} for an intersection scenario.
In Section~\ref{subsec:sensitivity}, we discuss a case in which the extreme curvature of the transformed cost $\runningcost_i(t; \linx; \dots)$ alluded to also in Section~\ref{subsec:soundness} which causes Algorithm~\ref{alg:flin_ilq_game} to converge very slowly.
In practice, however, this is not necessarily a serious problem. In Section~\ref{subsec:designing_costs}, we redesign this problematic cost function to depend explicitly upon $\linx$ rather than $x$ without changing the semantic character of equilibria.

\subsection{Improvements in solver stability}
\label{subsec:stability}

To showcase the benefits of our feedback linearization-based approach, we study the empirical sensitivity of solutions to the initial step size $\stepsize_0$ and trust region size $\trustregion$ hyperparameters from Section~\ref{subsec:main_alg}. 
We shall consider a three-player intersection example and compare the strategies identified by Algorithm~\ref{alg:flin_ilq_game} with those identified on the original dynamics, using the algorithm from \cite{fridovich2019efficient}.
Here, two cars, modeled with bicycle dynamics
\begin{align}
    \label{eqn:bicycle_dyn}
    \begin{bmatrix}
    \dot p_x\\
    \dot p_y\\
    \dot \theta
    \end{bmatrix}=\begin{bmatrix}
    v \cos\theta\\
    v \sin\theta\\
    (v / L) \tan \phi
    \end{bmatrix},~
    \begin{bmatrix}
    \dot v\\
    \dot \phi\\
    \dot a
    \end{bmatrix}=\begin{bmatrix}
    a\\
    \omega\\
    \kappa
    \end{bmatrix}
\end{align}
(with inter-axle distance $L$ and inputs $\omega$ controlling front wheel rate $\dot\phi$ and $\kappa$ controlling jerk), and a pedestrian modeled with dynamics \eqref{eqn:unicycle_dyn} navigate an intersection.
Like \eqref{eqn:unicycle_dyn}, bicycle dynamics \eqref{eqn:bicycle_dyn} are feedback linearizable in the outputs $(p_x, p_y)$.
We place quadratic penalties on each player's distance from the appropriate lane center and from a fixed goal location, as well as on the difference between speed $v$ and a fixed nominal speed $\bar v$.
Players are also penalized quadratically within a fixed distance of one another \footnote{For details concerning weighing of different cost terms we refer the reader to our github repository at \url{https://github.com/HJReachability/ilqgames}}.
% The examples we will be using will correspond to: (a) a three player intersection problem and, (b) a three player overtaking maneuver depicted in Fig.~\ref{fig:examples}. 

% \begin{figure}[htbp]
%     \centering
%     \includegraphics[width=\columnwidth]{example-image-a}
%     \caption{Caption.}
%     \label{fig:examples}
%     %\vspace{-.2cm}
% \end{figure}

In order to assess the quality of a trajectory ${\traj = (\linx, \auxinput_1, \dots, \auxinput_\numplayers)}$ generated by a particular algorithm, we define the \textit{similarity metric} to the desired trajectory to be:
\begin{align}
\label{eqn:quality_metric}
    q(\traj,\tilde\traj) := \max_{t\in[0, \horizon]} \|\linx(t) - \tilde\linx(t)\|_{2, (p_x, p_y)}.
\end{align}
Here, we take $\tilde\traj := (\tilde\linx, \tilde\auxinput_1, \dots, \tilde\auxinput_\numplayers)$ to be the equilibrium trajectory which that algorithm ideally converges to. The norm measures Euclidean distance only in the $(p_x, p_y)$ dimensions. 
%For a given $\traj$, the \textit{min} in this metric first finds the smallest distance between both trajectories, while the \textit{max} finds the maximimum over all minimum deviations.
Trajectories that diverge or converge to unreasonable solutions yield high values for $q$, while trajectories that closely match $\tilde\traj$ incur low values.

\begin{figure}[tbp]
    \centering
    \includegraphics[width=0.85\linewidth, trim=60 278 70 280, clip=true]{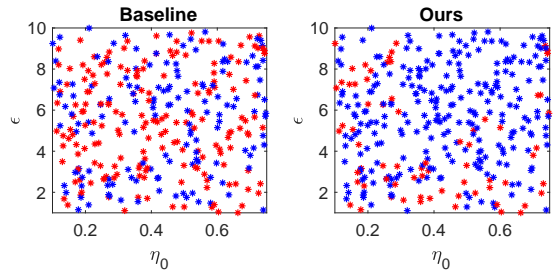}
    % \vspace{-0.1cm}
    \caption{Distribution of pairs $(\stepsize_0, \trustregion)$ colored by quality metric $q$. Pairs with low $q$ are colored blue, and high $q$ pairs are colored red. 
    }
    \label{fig:2d_distribution_intersection}
    \vspace{-.5cm}
\end{figure}

\begin{figure}[tbp]
    \centering
    \includegraphics[width=0.83\linewidth, trim=100 45 120 85, clip=true]{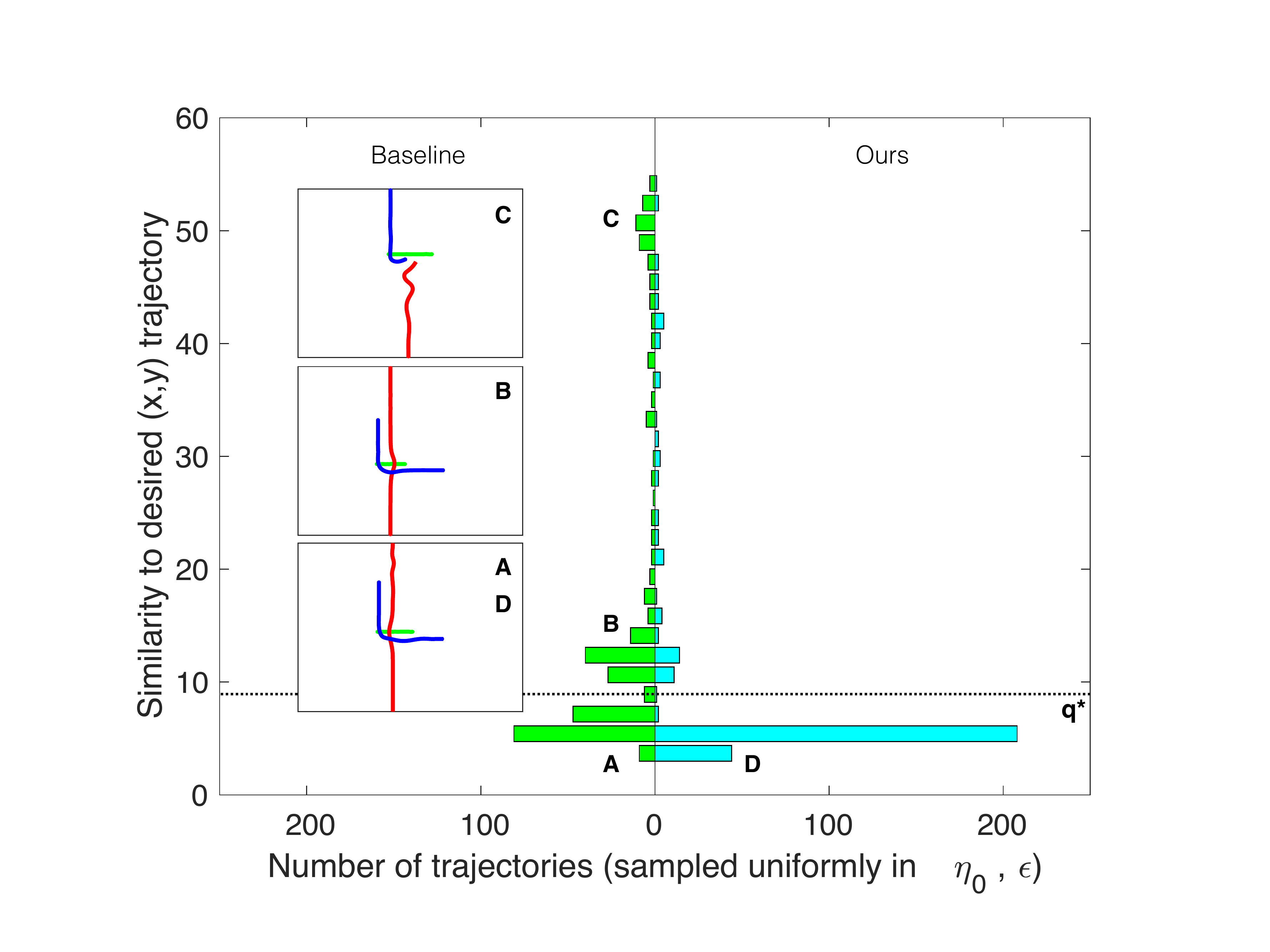}
    \caption{
    Comparison of the proposed algorithm with the state of the art \cite{fridovich2019efficient} for a three player intersection game. 
    Histograms (left, baseline; right, ours) show that our method is much more numerically stable and converges more frequently.
    % Distribution of $q(\stepsize_0;\trustregion)$ over 324 experiments for each solver. 
    % The histogram on the left (blue) represents $q$-values for the original solver of \cite{fridovich2019efficient}, while the histogram on the right (red) corresponds to Algorithm~\ref{alg:flin_ilq_game}.
    Insets labelled \{A, B, C, D\} show a typical trajectory for the associated bin. The dotted horizontal line shows threshold $q^*$, used to distinguish samples from Fig.~\ref{fig:2d_distribution_intersection}.
    }
    \label{fig:hist_intersection}
    \vspace{-.6cm}
\end{figure}

We fix the initial conditions and cost weights identically for both algorithms. Thus, any trajectory $\traj$ identified by the solver will solely be a function of the initial step size $\stepsize_0$ and trust region size $\trustregion$. Therefore, we will overload the penalty metric notation as $q(\stepsize_0; \trustregion)$. Given this metric we study the quality of solutions over the ranges $\stepsize_0 \in [0.1,0.75]$ and $\trustregion \in [1.0,10.0]$, and test $324$ uniformly sampled $(\stepsize_0, \trustregion)$ pairs. 

Fig.~\ref{fig:2d_distribution_intersection} displays the sampled pairs over the space of $\stepsize_0$ and $\trustregion$. For clarity, we set a \textit{success threshold} $q^*$ and color ``successful'' pairs with $q(\stepsize_0; \trustregion) \le q^*$ blue, and ``unsucessful'' pairs red. 
% \subsection{Results for Three Player Intersection}
% \label{label}
Fig.~\ref{fig:hist_intersection} shows histograms of solution quality $q$ for each algorithm, with a horizontal line denoting threshold $q^*$. 
We observe that solving the game using feedback linearization converges much more reliably than solving it for the original nonlinear system.
Moreover, for converged trajectories with low $q$-value, the average computation time was $0.3982 \pm 0.3122$ s (mean $\pm$ standard deviation) for our method and ${0.8744 \pm 0.9582~\textnormal{s}}$ for the baseline.

\subsection{Sensitivity to transformed cost landscape}
\label{subsec:sensitivity}

Unfortunately, these results do not generalize to all games. As per Section~\ref{subsec:soundness}, in some cases the cost landscape gets much more complicated when expressed in linearized system coordinates $\linx, \auxinput_i$. 
For example, a simple quadratic penalty on a single player's speed difference from nominal $\bar v$ in \eqref{eqn:unicycle_dyn} is nonconvex and non-smooth near the origin when expressed as a function of linearized system state $\linx$:
\begin{align}
    \label{eqn:v_cost_transformation}
    (v - \bar v)^2 \iff \left(\bar v - \sqrt{\dot p_x^2 + \dot p_y^2}\right)^2.
\end{align}
Consequences vary; the effect is negligible in the intersection example from Fig.~\ref{fig:hist_intersection}, but it is more significant in the roundabout example below in Section~\ref{subsec:designing_costs}, where cars must slow down before turning into the roundabout.

\subsection{Designing costs directly for the linearized system}
\label{subsec:designing_costs}

Fortunately, in practical settings of interest it is typically straightforward to design smooth, semantically equivalent costs explicitly as functions of the linearized system coordinates $\linx$.
For example, we can replace the nominal speed cost of \eqref{eqn:v_cost_transformation} with a time-varying quadratic penalty in that player's position $(p_x, p_y)$:
\begin{align}
    \label{eqn:route_progress_cost}
    (v - \bar v)^2 \implies \big(p_x(t) - \bar p_x(t)\big)^2 + \big(p_y(t) - \bar p_y(t)\big)^2,
\end{align}
where $\big(\bar p_x(\cdot), \bar p_y(\cdot)\big)$ defines the point on the lane center a distance $\bar v t$ from the initial condition.
% \subsection{Results for Three Player Overtaking}

\begin{figure}[tbp]
    \centering
    \includegraphics[width=0.83\columnwidth, trim=100 45 120 85, clip=true]{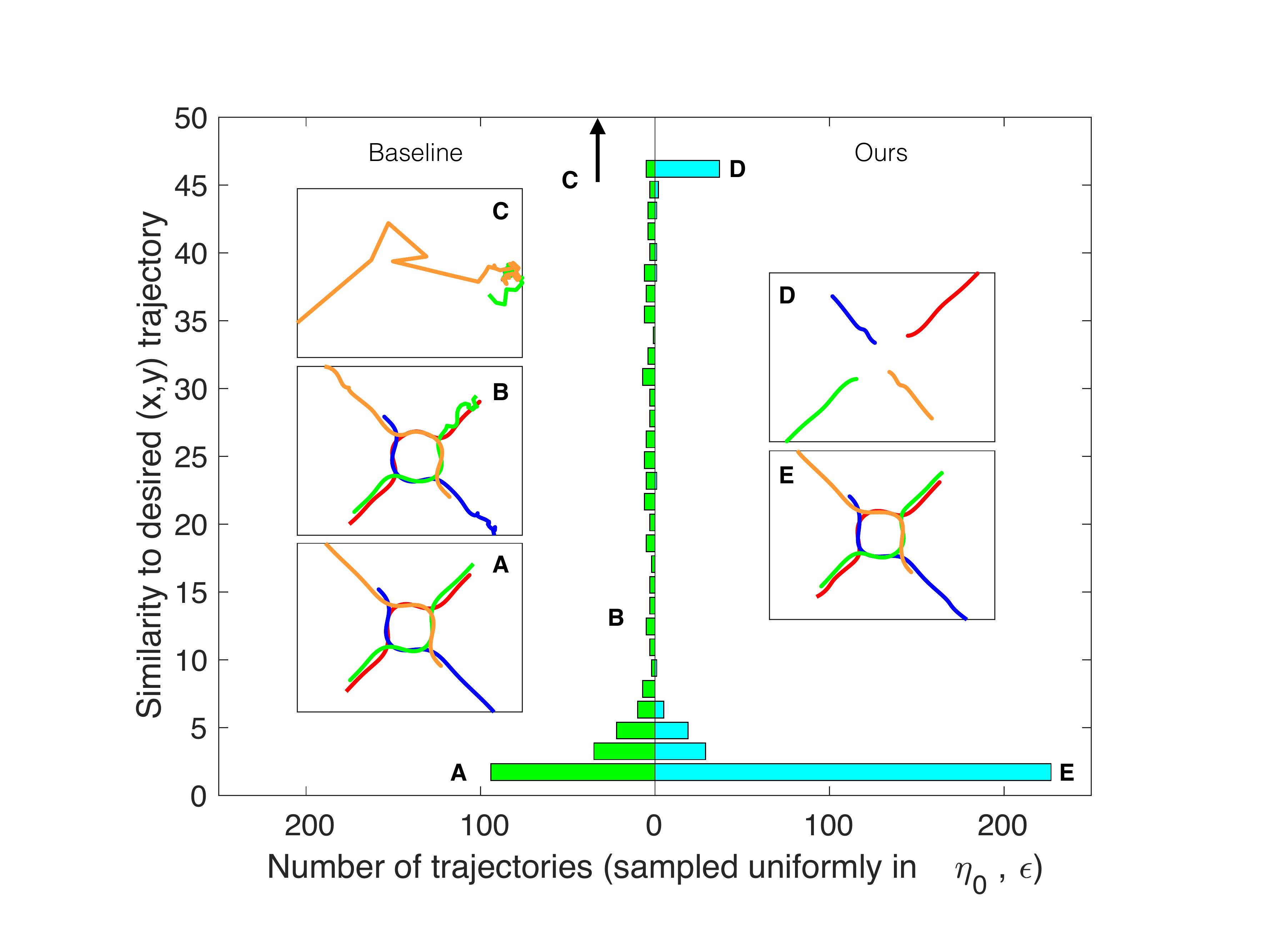}
    \caption{Comparison for a roundabout merging example with four cars. 
    %Histograms show the distribution of quality metric $q$ for the same 324 $(\stepsize_0, \trustregion)$ pairs in Section~\ref{subsec:stability}. Insets show representative solutions from different bins.
    }
    \label{fig:hist_roundabout}
    \vspace{-.2cm}
\end{figure}

\begin{figure}[tbp]
    \centering
    \includegraphics[width=0.83\columnwidth,trim=100 45 120 85, clip=true]{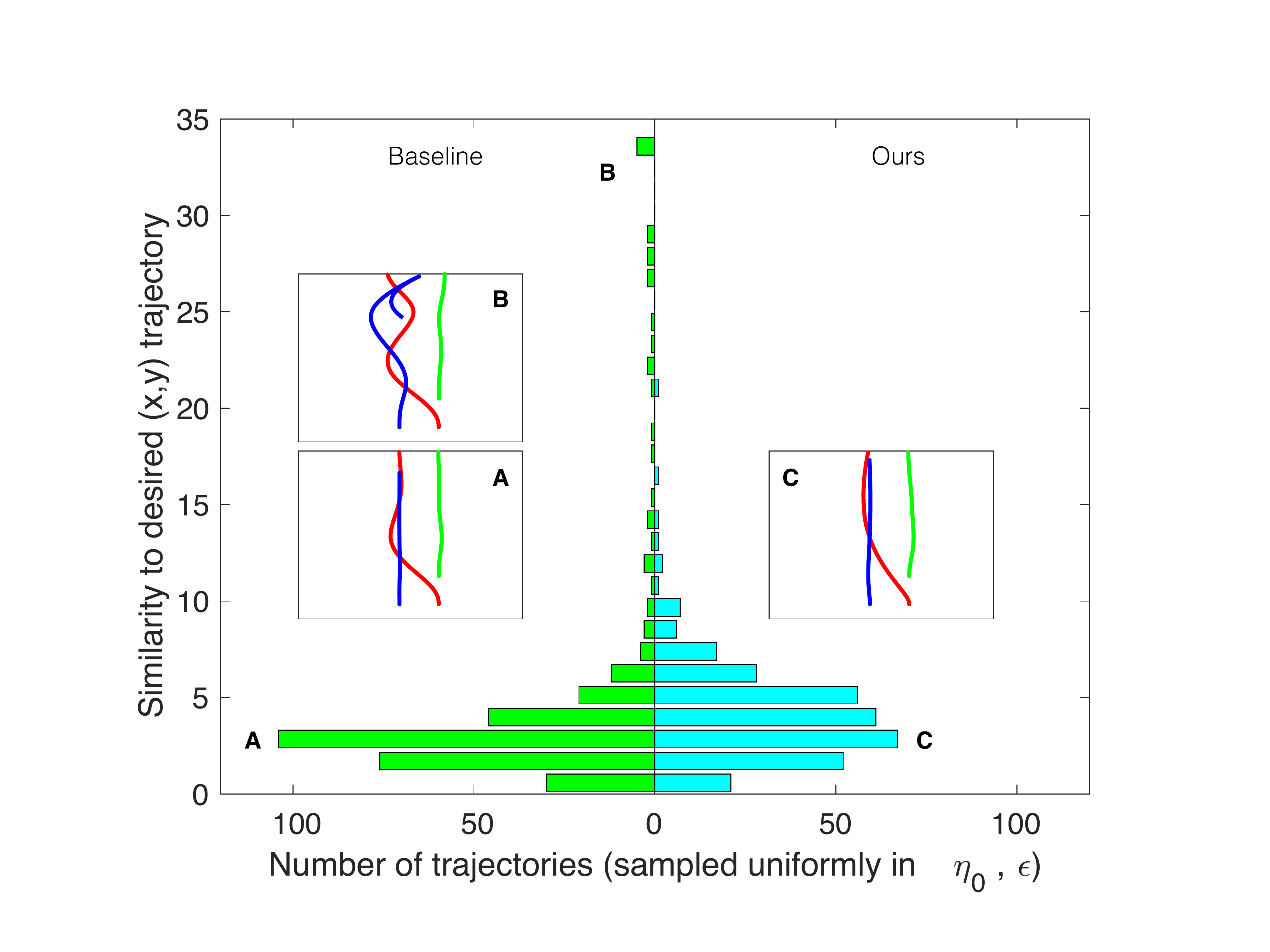}
    \caption{Comparison for a three vehicle high speed overtaking maneuver.}
    \label{fig:hist_overtaking}
    \vspace{-.6cm}
\end{figure}

We demonstrate the effectiveness of this substitution in two examples---merging into a roundabout, and overtaking a lead vehicle---in which the original cost \eqref{eqn:v_cost_transformation} led to instability in Algorithm~\ref{alg:flin_ilq_game}.
In both cases, we also use simple quadratic penalties for $\auxinput_i$ (rather than transforming $\|u_i\|^2$ into linearized coordinates), albeit with different weightings.
Results for the roundabout merging and overtaking examples are shown in Figures~\ref{fig:hist_roundabout} and \ref{fig:hist_overtaking}, respectively. From the 324 samples in each (drawn from expanded ranges $\stepsize_0 \in [0.1,1.0], \trustregion \in [1,50]$), we see that Algorithm~\ref{alg:flin_ilq_game} converged more frequently than the method of \cite{fridovich2019efficient}. Moreover, when successful, the average computational time in the roundabout example was $0.2797 \pm 0.1274$ s for our method and $0.4244 \pm 0.5259$ s for the baseline. Runtimes for the overtaking example were $0.5112 \pm 0.3228$ s (ours) and $0.4417 \pm 0.4142$ s (baseline). Observe how runtimes for our approach cluster more tightly around the mean, indicating a more reliable convergence rate.

% To further investigate the behavior of the modified solver, we also display the sampled pairs over the space of $\stepsize_0$ and $\epsilon$. To aid in the visualization we partition and color the pairs according to a \textit{success threshold} $q^*$. Pairs which result in trajectories for which $q(\stepsize_0,\epsilon) \leq q^*$ are marked as successes and denoted in blue, while failures are denoted in red.  

% \begin{figure}[htbp]
%     \centering
%     \includegraphics[width=\columnwidth]{example-image-c}
%     \caption{Caption.}
%     \label{fig:2d_distribution_overtaking}
%     %\vspace{-.2cm}
% \end{figure}

% % \todoin{
% In both categories, show some nice quantitative plots of performance for 1-2 example problems:
% \begin{enumerate}
% \item runs faster, esp. from zero initialization, which is important for real planning problems when new agents appear. Should also have hardware example here.
% \item solver is less sensitive to step size $\eta$
% \end{enumerate}
% }

\section{Conclusion}
\label{sec:conclusion}

We have presented a novel algorithm for identifying local equilibria in differential games with feedback linearizable dynamics.
Our method works by repeatedly solving LQ games in the linearized system coordinates, rather than in the original system coordinates.
By working with the linearized system, our algorithm becomes less sensitive to parameters such as initial step size and trust region size, which often leads it to converge faster.
Our method is fully general, i.e. any cost expressed in terms of nonlinear system coordinates may also be expressed in terms of linearized coordinates.
However, in some cases transforming costs in this way makes the cost landscape extremely complicated.
In such cases, it is often possible to design semantically equivalent replacement costs directly in the linearized coordinates.
We test our method in a variety of competitive traffic scenarios. Using appropriately redesigned costs when necessary, our experiments confirm the computational stability and efficiency of our approach. 

% \todoin{
% We should definitely address the randomization of alpha+trregion but not of the weights. i.e. maybe there is a set of weights for which our results are different.
% }

% \section*{Acknowledgments}

% \bibliographystyle{plainnat}
%\bibliography{references}
\balance
\printbibliography

@book{basar1999dynamic,
  title={Dynamic Noncooperative Game Theory},
  author={Ba\c{s}ar, Tamer and Olsder, Geert Jan},
%  volume={23},
  year={1999},
  publisher={SIAM}
}

@article{dreves2019best,
  title={A best-response approach for equilibrium selection in two-player generalized Nash equilibrium problems},
  author={Dreves, Axel},
  journal={Optimization},
  volume={68},
  number={12},
  pages={2269--2295},
  year={2019},
  publisher={Taylor \& Francis}
}

@article{dreves2018generalized,
  title={A generalized Nash equilibrium approach for optimal control problems of autonomous cars},
  author={Dreves, Axel and Gerdts, Matthias},
  journal={Optimal Control Applications and Methods},
  volume={39},
  number={1},
  pages={326--342},
  year={2018},
  publisher={Wiley Online Library}
}

@inproceedings{williams2018best,
  title={Best response model predictive control for agile interactions between autonomous ground vehicles},
  author={Williams, Grady and Goldfain, Brian and Drews, Paul and Rehg, James M and Theodorou, Evangelos A},
  booktitle={2018 IEEE International Conference on Robotics and Automation (ICRA)},
  pages={2403--2410},
  year={2018},
  organization={IEEE}
}

@article{fridovich2019efficient,
	title = {Efficient Iterative Linear-Quadratic Approximations for Nonlinear Multi-Player General-Sum Differential Games},
	journal = {arXiv preprint arXiv:1909.04694},
	author = {Fridovich-Keil, David and Ratner, Ellis and Dragan, Anca D and Tomlin, Claire J},
	year={2019}
}

@incollection{richter2016polynomial,
  title={Polynomial trajectory planning for aggressive quadrotor flight in dense indoor environments},
  author={Richter, Charles and Bry, Adam and Roy, Nicholas},
  booktitle={Robotics Research},
  pages={649--666},
  year={2016},
  publisher={Springer}
}

@inproceedings{mellinger2011minimum,
  title={Minimum snap trajectory generation and control for quadrotors},
  author={Mellinger, Daniel and Kumar, Vijay},
  booktitle={International Conference on Robotics and Automation},
  pages={2520--2525},
  year={2011},
  organization={IEEE}
}

@inproceedings{rouchon1993flatness,
  title={Flatness, motion planning and trailer systems},
  author={Rouchon, Pierre and Fliess, Michel and L{\'e}vine, Jean and Martin, Philippe},
  booktitle={Conference on Decision and Control (CDC)},
  volume={3},
  pages={2700--2700},
  year={1993},
  organization={IEEE}
}

@article{murray1993nonholonomic,
  title={Nonholonomic motion planning: Steering using sinusoids},
  author={Murray, Richard M and Sastry, Sosale Shankara},
  journal={Transactions on Automatic Control},
  volume={38},
  number={5},
  pages={700--716},
  year={1993},
  publisher={IEEE}
}

@inproceedings{jonsson2011scaling,
  title={Scaling up multiagent planning: A best-response approach},
  author={Jonsson, Anders and Rovatsos, Michael},
  booktitle={International Conference on Automated Planning and Scheduling},
  year={2011}
}

@book{sastry1999nonlinear,
  title={Nonlinear Systems: Analysis, Stability, and Control},
  author={Sastry, Shankar},
  publisher={Springer},
  year={1999}
}

@techreport{isaacs1951games,
  title={Games of pursuit},
  author={Isaacs, Rufus},
  year={1951},
  institution={Rand Corporation}
}

@inproceedings{mukai2000sequential,
  title={Sequential linear quadratic method for differential games},
  author={Mukai, H and Tanikawa, A and Tunay, I and Katz, IN and Sch\"{a}ttler, H and Rinaldi, P and Ozcan, IA and Wang, GJ and Yang, L and Sawada, Y},
  booktitle={Proc. 2nd DARPA-JFACC Symposium on Advances in Enterprise Control},
  pages={159--168},
  year={2000},
  organization={Citeseer}
}

@article{starr1969nonzero,
  title={Nonzero-sum differential games},
  author={Starr, Alan Wilbor and Ho, Yu-Chi},
  journal={Journal of Optimization Theory and Applications},
  volume={3},
  number={3},
  pages={184--206},
  year={1969},
  publisher={Springer}
}

@article{starr1969further,
  title={Further properties of nonzero-sum differential games},
  author={Starr, AW and Ho, Yu-Chi},
  journal={Journal of Optimization Theory and Applications},
  volume={3},
  number={4},
  pages={207--219},
  year={1969},
  publisher={Springer}
}

@incollection{wang2019game,
  title={Game Theoretic Motion Planning for Multi-robot Racing},
  author={Wang, Zijian and Spica, Riccardo and Schwager, Mac},
  booktitle={Distributed Autonomous Robotic Systems},
  pages={225--238},
  year={2019},
  publisher={Springer}
}

@book{bertsekas1996neuro,
  title={Neuro-dynamic programming},
  author={Bertsekas, Dimitri P and Tsitsiklis, John N},
%  volume={5},
  year={1996},
  publisher={Athena Scientific Belmont, MA}
}

@article{jacobson1970differential,
  title={Differential dynamic programming},
  author={Jacobson, David H and Mayne, David Q},
  year={1970},
  publisher={North-Holland}
}

@inproceedings{li2004iterative,
  title={Iterative linear quadratic regulator design for nonlinear biological movement systems.},
  author={Li, Weiwei and Todorov, Emanuel},
  booktitle={ICINCO},
  pages={222--229},
  year={2004}
}

@inproceedings{van2014iterated,
  title={Iterated LQR smoothing for locally-optimal feedback control of systems with non-linear dynamics and non-quadratic cost},
  author={van den Berg, Jur},
  booktitle={American Control Conference (ACC)},
  pages={1912--1918},
  year={2014},
  organization={IEEE}
}

@inproceedings{wangmingyu2019game,
  title={Game Theoretic Planning for Self-Driving Cars in Competitive Scenarios},
  author={Wang, Mingyu and Wang, Zijian and Talbot, John and Gerdes, J Christian and Schwager, Mac},
  booktitle={Robotics: Science \& Systems},
  year={2019}
}

@inproceedings{tassa2014control,
  title={Control-limited differential dynamic programming},
  author={Tassa, Yuval and Mansard, Nicolas and Todorov, Emo},
  booktitle={International Conference on Robotics and Automation (ICRA)},
  pages={1168--1175},
  year={2014},
  organization={IEEE}
}

@inproceedings{sadigh2016planning,
  title={Planning for autonomous cars that leverage effects on human actions.},
  author={Sadigh, Dorsa and Sastry, Shankar and Seshia, Sanjit A and Dragan, Anca D},
  booktitle={Robotics: Science \& Systems},
%  volume={2},
  year={2016},
  organization={Ann Arbor, MI, USA}
}

\end{document}